\documentstyle[prd,aps]{revtex}
\begin{document}

\draft

\font\large=cmbx10 at 12 pt  
\newcount\equationno      \equationno=0
\newtoks\chapterno \xdef\chapterno{}
\def\eqn{\eqno\eqname}
\def\eqname#1{\global \advance \equationno by 1 \relax
\xdef#1{{\noexpand{\rm}(\chapterno\number\equationno)}}#1}

\def\la{\mathrel{\mathchoice 
{\vcenter{\offinterlineskip\halign{\hfil
$\displaystyle##$\hfil\cr<\cr\sim\cr}}}
{\vcenter{\offinterlineskip
\halign{\hfil$\textstyle##$\hfil\cr<\cr\sim\cr}}}
{\vcenter{\offinterlineskip
\halign{\hfil$\scriptstyle##$\hfil\cr<\cr\sim\cr}}}
{\vcenter{\offinterlineskip
\halign{\hfil$\scriptscriptstyle##$\hfil\cr<\cr\sim\cr}}}}}

\def\ga{\mathrel{\mathchoice 
{\vcenter{\offinterlineskip\halign{\hfil
$\displaystyle##$\hfil\cr>\cr\sim\cr}}}
{\vcenter{\offinterlineskip
\halign{\hfil$\textstyle##$\hfil\cr>\cr\sim\cr}}}
{\vcenter{\offinterlineskip
\halign{\hfil$\scriptstyle##$\hfil\cr>\cr\sim\cr}}}
{\vcenter{\offinterlineskip
\halign{\hfil$\scriptscriptstyle##$\hfil\cr>\cr\sim\cr}}}}}
\def\s{\smallskip}
\def\n{\noindent}

\def\x{{\bf x}}
\def\y{{\bf y}}
\def\R{{\cal R}}

\reversemarginpar


\title{Radiation from a charged particle and radiation reaction -- revisited}
\author{Abhinav Gupta\thanks{abh@ducos.ernet.in}}
\address{Department of Physics and Astrophysics, Delhi
	University, India}
\author{T. Padmanabhan\thanks{paddy@iucaa.ernet.in}}
\address{IUCAA, Post Bag 4, Ganeshkhind, Pune 411 007, India.}
\maketitle

\begin{abstract}

We study the electromagnetic fields of an arbitrarily moving charged 
particle and the radiation reaction on the charged particle using a novel
approach. We first show that the fields of an {\it arbitrarily} moving 
charged particle in an inertial frame can be related in a simple manner
to the fields of a {\it uniformly accelerated} charged particle in 
its rest frame. Since the latter field is static and easily obtainable,
it is possible to derive the fields of an arbitrarily moving 
charged particle by a coordinate transformation. More importantly,
this formalism allows us to calculate the self-force on a charged 
particle in a remarkably simple manner. We show that the original 
expression for this force, obtained by Dirac, can be rederived with 
much less computation and in an intuitively simple manner using our formalism.
\end{abstract}
\pacs{PACS number(s): }

\section{\bf Motivation}

The field of a charged particle at rest in an inertial frame is a static 
Coulomb field which falls as $(1/r^2)$ in the standard spherical coordinate system. The field of a charge, moving with uniform velocity, can 
be obtained by Lorentz transforming the Coulomb field; this field 
also falls as inverse square of the distance. The situation changes dramatically
for a charged particle which is moving with non zero acceleration. The field
now has a piece which falls only as $(1/r)$, usually called the radiation
field. For a field which decreases as $(1/r)$, the energy flux varies
as $(1/r^2)$ implying that the same amount of energy flows through 
spheres of different radii at sufficiently large distances from the 
charge. Because of this reason, the radiation fields acquire a life 
of their own and the entire phenomena of electromagnetic radiation hinges
on this feature. Due to the continuous transfer of energy from the 
charged particle to large distances, there will be a damping force acting 
on the charged particle which is usually called the radiation reaction 
force. The derivation of radiation reaction force is conceptually and 
operationally quite complicated and the expression --- obtained originally 
by Dirac (see [4])--- has no simple intuitive description. 

We analyse these issues from a novel point of view in this paper which throws
light on the conceptual and mathematical issues involved in this problem.
The analysis is motivated by the following issue: Maxwell's equations 
are not only Lorentz invariant but can also be written in a generally 
covariant manner. Given a charged particle moving in some arbitrary trajectory,
it is always possible to construct a proper coordinate system for such a 
charged particle. In such a coordinate system, the charge will be at rest 
for all times  but the background metric will be non Minkowskian and 
-- in general -- time dependent. The Maxwell's equations in this coordinate
system will correspond to that of a {\it stationary} charge located 
in a non trivial (and in general time dependent) metric.
The solution to Maxwell's equation in this frame receives time dependent contributions
{\it not} because of the motion of charged particles but because of the non trivial 
nature of the {\it background} metric. But we know that, for internal consistency,
these solutions should transform to the standard solutions describing the 
field of an arbitrarily moving charged particle when we go over to the 
inertial frame! This is remarkable since the time dependence and nontriviality
of the {\it background metric} have to translate to the correct spatial and  time dependence of the {\it radiation field}. Further, the charged particle has to 
feel the radiation reaction force in the non inertial frame, 
even though it is at rest, due to the non triviality of the background metric.
It is not intuitively obvious how these features come about and it is 
important to understand how the physics in the non inertial frame of the 
charged particle operates. 

We shall explore in this paper both the issues raised in the above paragraph.
The key feature which emerges from our analysis is the following. The 
structure of the Maxwell's equations dictate that the static field of a 
{\it uniformly accelerated} charged particle in its rest frame can be related to the field of an {\it arbitrarily moving} charged particle in the inertial
frame. This connection also carries over to the computation of the self-force.
It turns out that the radiation reaction force has a simple geometrical origin in the uniformly accelerating frame in which the charged particle is 
instantaneously at rest. The force arises due to the deviation of the
trajectory of the charged particle from that of uniform acceleration and hence
is proportional to the derivative of the acceleration. We shall now spell out
the details of the approach we plan to follow in this paper.
\section{\bf The formalism} 

Consider the electromagnetic field of a charge moving with a uniform velocity in an inertial frame $S$. Since Maxwell's equations are
Lorentz covariant, the most natural way to calculate the field in $S$ is to find
the field in the charge's rest frame $S'$  and  transform back to $S$.
Let us next consider  the problem of calculating the electromagnetic field of a charge which is moving {\it arbitrarily}. The conventional method (see e.g. \cite{ll72}) is to calculate the Leinard - Weichert potential and to differentiate it to obtain the field. However, we will  show that it is possible to approach the problem differently along the following lines:

Consider a charge moving with an arbitrary velocity and acceleration in an inertial
frame $S$.
 In the Lorentz gauge, Maxwell's equations can be written in terms of the vector potential $A^{\mu}$ and the current $j^{\mu}$ as: 
\begin{equation}
\Box 
A^{\mu}=4{\pi}j^{\mu} \label{eqn:one}
\end{equation}
where~$\Box =\partial_{\mu}\partial^{\mu}$ .
It follows that:
\begin{equation}
\Box F^{\mu \nu}=4{\pi}(\partial^{\mu}j^{\nu}-
\partial^{\nu}j^{\mu})\label{eqn:two}
\end{equation}
Because of the characteristics of the $\Box$ operator,  the fields at an event $P$ can only depend on 
the trajectory  of the charge at the retarded event $O$, which is the point of intersection of the backward light cone drawn from $P$, and the worldline of the charge $z^{\mu}=z^{\mu}({\tau})$. Since $j^{\mu}$ is linear in four velocity, the quantity $\partial_{\mu}j_{\nu}$, in the  the right hand side of (\ref{eqn:two}) can at most depend on ${\ddot z}^{\mu}(\tau)$. Therefore, the fields at $P$ can at most depend on the second derivatives at the retarted position of the charge at $O$ - i.e.,  at most on the retarded acceleration of the charge.
Suppose we now change the trajectory of the charge to that of a uniformly
accelerated one without changing the values of the velocity and
acceleration at the retarded event $O$. The field at $P$, since it depends only
on the velocity and acceleration at $O$ will still remain the same. It follows
that, if we know the field at $P$ due to a uniformly accelerated charge
with a given acceleration and  velocity at $O$, then we can obtain
the field due to a general trajectory.

Thus the problem reduces to that of calculating the field of a uniformly
accelerated charged particle. This is most easily done by using the fact
that Maxwell's equations can be written in a generally covariant manner.
Solving the Maxwell's equations  in the noninertial, rest frame of charge
and transforming the field to the inertial frame, we can obtain the
field of a uniformly accelerated particle. Using the argument outlined above,
we can then find the field of a charged particle moving in an arbitrary trajectory.
To illustrate the power of this technique, we shall directly calculate the field
for arbitrary, {\it rectilinear} motion. (The general case is a straightforward extension,
and is treated in Appendix C).

The real power of this formalism, however, lies in calculating the field in the
{\it infinitesimal neighbourhood}  of the accelerating charge. The general expression
for the field in the neighbourhood  of an accelerating charge, found by Dirac, is a fairly involved expression, and good deal of labour is required to
compute it.  Our formalism 
involves computing it in the instantaneous coaccelerating frame of the charge,
in which  the first and second derivatives of the position
of the charge vanish. The only dynamical contribution to the near field comes from the the third derivative, which --- as we shall see --- leads to the radiation reaction term.
This, along with the static terms, neatly transforms into expression obtained by Dirac in
the inertial frame. In addition, a novel interpretation for radiation reaction
emerges in the accelerated frame.

The rest of the paper is organised as follows: In section 3, we obtain the electromagnetic field of a uniformly accelerated charge. This is done by solving Maxwell's equations in the rest frame of the charged particle (which is a noninertial frame) and transforming to the inertial frame. In section 4, we use this result to obtain the field of an arbitrarily moving charged particle. This result is obtained by the procedure outlined above. Section 5 uses the same formalism to obtain the field in the neighbourhood of the charged particle, thereby obtaining the radiation reaction term. The last section summarises the results of the paper.
\newpage

\section{Fields due to a charge at rest in a uniformly accelerated frame}
\subsection{ The coordinate transformation}

Since the key idea involves working with a uniformly accelerated frame,
we shall review the coordinate transformation connecting the Minkowski
frame to the Rindler frame and collect together the necessary formulas. 

Consider a charge moving with uniform acceleration along the z-axis of an inertial frame $S$ with the  coordinate system $(t,x,y,z)$. The trajectory of the charge is given by:
\begin{equation}
t=\frac {1}{g}\sinh(g\tau) ~;~
z=\frac {1}{g}\cosh(g\tau) \label{eqn:three}
\end{equation}
where $g$  is the proper acceleration of the charge, and ${\tau}$ is its proper time. The world line, 

\begin{equation}
z^2-t^2=(\frac {1}{g})^2 \label{eqn:four}
\end{equation}
is a hyperbola.
Referring to figure 1, one can see that this  charge can influence  regions $A$ and $B$ of spacetime, which lie along the forward light cone of
the charge's trajectory but not the regions $C$ and $D$. 
Let us now fix a proper, Fermi-Walker transported coordinate system $({\tau},{\zeta},x,y)$ to the accelerating
charge and call it frame $U$. Separate transformations are defined from $S$ to $U$ for regions $A$ and $B$. 
In region $A$, we take 
\begin{equation}
t={\frac {\sqrt {2g\zeta}}{g}}\sinh(g\tau)~;~
 z={\frac {\sqrt {2g\zeta}}{g}}\cosh(g\tau)~;~\zeta>0.\label{eqn:five}
\end{equation}
and in region $B$ we take
\begin{equation}
t={\frac {\sqrt {-2g\zeta}}{g}}\cosh(g\tau)~;~
z={\frac {\sqrt {-2g\zeta}}{g}}\sinh(g\tau)~;~\zeta<0\label{eqn:six}
\end{equation}
$x$ and $y$ are mapped to themselves. The spacetime interval, both in 
 regions $A$ and $B$ is :
\begin{equation}
ds^2~=~2g{\zeta}~d{\tau}^2~
-{\frac {{d\zeta}^2}{2g\zeta}}~-d{\rho}^2~-{{\rho}^2}d{\phi}^2 \label{eqn:seven}
\end{equation}
where ${\rho}={\sqrt{x^2+y^2}}\ {\rm and}\  {\phi}={\tan^{-1}}(y/x)$.
The range $[{\zeta}>0~;~ -{\infty}< {\tau}<~+{\infty}]$ covers region $A$, and $[{\zeta}<0~;~ -{\infty}<{\tau}<~+{\infty}]$ covers region $B$ .
 In these coordinates, the charge is at rest, at  ${\zeta}_0=(1/2g)$.

Since the metric is same for the transformations defined by equations (\ref{eqn:five}) and (\ref{eqn:six}) , we can solve Maxwell's equations in the background metric of (\ref{eqn:seven}) and transform separately in regions $A$ and $B$ to get the fields in the frame $S$.

\subsection{ The fields in the accelerated frame}

Let us next obtain the solutions to the Maxwell's equations in the 
noninertial Rindler frame.
The generally covariant form of Maxwell's equations are :
\begin{equation}
{\frac{1}{\sqrt{-g}}}{\partial_\mu}({\sqrt{-g}}F^{\mu\nu})=4{\pi}j^{\nu}\label{eqn:eight}
\end{equation}
 with
\begin{equation}
F_{\mu\nu}~=~{\partial_\mu}~A_{\nu}~-~{\partial_\nu}~A_{\mu}\label{eqn:thirteen}
\end{equation}
The current is:
\begin{equation}
j^{\mu}={\frac{e}{\sqrt{-g}}}{\delta^{3}}({\bf x } -{\bf x}_0)~{\frac{dx^\mu}{dx^0}}
\label{eqn:ten}
\end{equation}
where ${\bf x}_0 = (\zeta_0, 0, 0)$ is coordinate of the charged particle in the accelerated frame and the Dirac delta function is defined to be
\begin{equation}
{\delta^3}({\bf x} - {\bf x}_0) ={\delta(\zeta-\zeta_0)}{\delta(\rho)}{\delta(\phi)}~;~{\int}{\delta^3}({\bf \vec x})~d{\zeta}~d{\rho}~d{\phi}=1\label{eqn:eleven}
\end{equation}
for a point charge at $\zeta=\zeta_0$.
Since the charge is at rest, $j^i=0~{\rm for} ~i=1,2,3~~{\rm and}~~j^0\neq 0$.
Correspondingly,we can take $\it A_i$=0 with all time derivatives vanishing.  Hence the only relevant components of the field tensor are
\begin{equation}
F^{\zeta 0} = -F_{\zeta 0} = -\partial_{\zeta} A_0; \qquad F^{\rho 0}~=~{-}{\frac{1}{2g\zeta}}\left(\partial_{\rho}~A_{0}\right)\label{eqn:fifteen}
\end{equation}
Expressing the field tensor in terms of the potential, we get the equation satisfied by $A_0$:
\begin{equation}
\rho{\frac{\partial^2A_0} {\partial \zeta^2}}+
{\frac{1}{2g \zeta}}
{\frac {\partial}{\partial\rho}}
(\rho
{\frac{\partial A_0}{\partial \rho}})
~=~-4\pi{e}{~\delta}^3({\bf x} - {\bf x}_0)\label{eqn:sixteen}
\end{equation}
This equation has a simple, closed form solution which can be obtained by direct integration of (\ref{eqn:sixteen}) for ${\bf x} \not = {\bf x}_0$ and matching the boundary condition at ${\bf x}  = {\bf x}_0$
\begin{equation}
A_0=ge{\frac{\zeta+{\zeta}_0+(1/2)g{\rho}^2}{\sqrt{(\zeta-{\zeta}_0+(1/2)g{\rho}^2)^2~+~(2g{\rho}^2{\zeta}_0)}}} = ge {\frac{\zeta+{\zeta}_0+(1/2)g{\rho}^2}{\sqrt{(\zeta + {\zeta}_0+(1/2)g{\rho}^2)^2  - 4 \zeta_0 \zeta }}}\label{eqn:seventeen}
\end{equation}
(An alternative  derivation of this  solution is  given in appendix A.)
Also, as mentioned earlier, $A_i$=0 implying that there are  no magnetic fields.

Let us next compute the electric field corresponding to this potential. In an inertial frame, $ {F^i}_0, {F^0}_i, F^{i0}$, can all be interpreted as defining the electric field (apart from difference in signs). However, in the metric defined by equation (\ref{eqn:seven}), these components have  different spatial dependence due to raising and lowering by $\it g_{\mu\nu}$, which is not constant.  So, in order to define the  {\it physical}  electric field, we go back to the basic definition of  electric field as the ``electromagnetic force per unit charge, experienced by a charge at rest''.
 The contravariant electromagnetic force vector is:
\begin{eqnarray}
f^{~\mu} = e {{F^{\mu}}_{\nu}}{\frac{dx^{\nu}}{ds}}\nonumber
\end{eqnarray}
  which for a charge at rest gives the electric field
\begin{equation} 
E^i \equiv {{F^{i}}_{0}}{\frac{dx^{0}}{ds}}
=\frac{{F^i}_0}{\sqrt{{g}_{00}}}\label{eqn:eighteen}
\end{equation}
Using this in  equations  (\ref{eqn:fifteen}), we get the electric field components :
\begin{equation}
E^{\zeta}=\frac{(2ge{\zeta}_0){\sqrt{2g{\zeta}}}([\zeta- \zeta_0-(1/2)g{\rho}^2])}{{\xi}^3}; \quad 
E^{\rho}=\frac{(2ge{\zeta}_0){\rho}{\sqrt{2g{\zeta}}}}{{\xi}^3}; \quad
E^{\phi}=0 \label{eqn:nineteen}
\end{equation}
 where
\begin{equation}
\xi\equiv\sqrt{(\zeta-\zeta_0+(1/2)g{\rho}^2)^2~+~2g{\rho}^2{\zeta}_0}
\end{equation}

There are some interesting features which are worth noting about this field. To simplify the analysis let us transform from the coordinates $(\tau, \zeta, \rho, \varphi)$ to $(\tau, Z, \rho, \varphi)$ where $\zeta = (gZ^2/2)$. The metric in region $A$ is now
\begin{equation}
ds^2 = g^2Z^2d \tau^2 - (dZ^2 + d\rho^2 + \rho^2d\varphi^2)
\end{equation}
The Z-component of the electric field in this coordinate system is
\begin{eqnarray}
E^Z &=& \frac{4e} {g^2} \frac{[Z^2 - \rho^2-g^{-2}]} {[(Z^2 + \rho^2 - g^{-2})^2 + 4 g^{-2} \rho^2]^{3/2}}\nonumber \\
&=& \frac{4e} {g^2} \frac{[Z^2 - \rho^2-g^{-2}]} {[(Z^2 + \rho^2 + g^{-2})^2 - 4 g^{-2} Z^{2}]^{3/2}}\label{eqn:nineteena}
\end{eqnarray}
In this coordinate system, the (apparent) event horizon is at $Z=0$. On this surface, the electric field is along $Z$ axis and has the value
\begin{equation}
E^Z ({\rm at} \quad Z=0) = - \frac {4e} {g^2} \frac {1} {(\rho^2 + g^{-2})^2}
\end{equation}
This is finite and is equivalent to having a charge density
\begin{equation}
\sigma({\rm at} \quad Z=0) = +\frac {E^z} {4 \pi} = -\frac {e} {\pi g^2} \frac {1} {(\rho^2 + g^{-2})^2}
\end{equation}
at the apparent horizon (This point was earlier noted in reference [2]). {\it Note that this result is coordinate dependent}. The field $E^{\zeta}$, in the coordinates $(\tau, \zeta, \rho, \varphi)$, {\it vanishes} at the horizon. In these coordinates, there is no charge density on the horizon.

If we shift the origin of Z-axis by introducing the coordinate $\bar Z = Z - g^{-1}$, then the metric becomes
\begin{equation}
ds^2 = (1 + g \bar Z)^2 d\tau^2 - (dZ^2 + d \rho^2 + \rho^2 d \varphi^2)
\end{equation} and the electric field becomes
\begin{equation}
E^{\bar z} =\frac {e\bar Z } {r^3} (1 + \frac {1} {2} \frac {gr^2} {\bar Z}) ( 1 + g \bar Z + \frac {1} {4} g^2r^2)^{-3/2}
\end{equation}
\begin{equation}
E^{\rho} = - \frac {e \rho} {r^3} ( 1 + \frac {1} {2} \frac {gr^2} {\bar Z})( 1 + g \bar Z )  ( 1 + g \bar Z + \frac {1} {4} g^2r^2)^{-3/2} ( 1 + g \bar Z - \frac {g} {2} \frac {r^2} {\bar Z} )^{-1}
\end{equation}
with $E^{\varphi} = 0$. In this form, it is clear that field is the usual coulomb field for $g \bar Z \ll 1, \quad gr \ll 1$.
The behaviour of the field near the charge, compared to its
form near the apparent horizon clearly shows the distorting effect of the background line element.

We shall now use this result to obtain the fields of an arbitrarily moving change. 
\section{Field of a charge moving rectilinearly, with arbitrary velocity and acceleration.}
\subsection{ The coordinate transformation}
We shall  calculate the field due to a rectilinearly moving charge using the approach described in section 2. Let this charge move along the $ z$-axis of the inertial frame $S$. We are interested in the  field at event  $P$ with coordinates $\it (t,z,{\rho},{\phi})$. The retarded event is $O$ with coordinates $(t_0,z_0,0,0)$. At $O$, let $v_{ret}$ be the velocity
 of the charge and $a_{ret}$ be its acceleration. Then, the proper acceleration is:
\begin{equation}
g=\sqrt{-a^{\mu}a_{\mu}}=a_{ret}{\gamma}^3, \label{eqn:twenty}
\end{equation}
where $\gamma=(1- v_{ret}^2)^{-1/2}$. We construct a comoving, uniformly accelerating observer with an attached coordinate frame $M$ with coordinates $\it (\tau,\zeta,\rho,\phi)$ such that the   origin of $M$ coincides 
with the world line  of the charge upto $\it {v}^\mu$ and $\it {a}^\mu$ at the event $O$. 
So, at $O$, in the frame $M$, the charge is instantaneously at rest without acceleration.  With this construction, the constant, proper,  acceleration of $M$ is $g$, as defined by equation (\ref{eqn:twenty}). 

The coordinate transformations from $S$ to $M$ are different in the region $A$ and $B$.  In region $A$, $(\zeta > 0)$
\begin{equation}
t=t_0-\frac{\gamma v_{ret}}{g}+{\frac{\sqrt{2g\zeta}}{g}}{\sinh(g\tau)};\qquad
z=z_0-\frac{\gamma}{g}+{\frac{\sqrt{2g\zeta}}{g}}{\cosh(g\tau)}\label{eqn:twentyone}
\end{equation}
while in region $B$, $(\zeta < 0 )$:
\begin{equation}
t=t_0-\frac{\gamma v_{ret}}{g}+{\frac{\sqrt{-2g\zeta}}{g}}{\cosh(g\tau)}; \qquad
z=z_0-\frac{\gamma}{g}+{\frac{\sqrt{-2g\zeta}}{g}}{\sinh(g\tau)}
\label{eqn:twentytwo} \end{equation}
The constants $(\it t_0 -\gamma v_{ret}/g)$ and $(z_0 - \gamma /g)$ ensure the condition that the charge is at rest and with zero  acceleration at $\zeta_0=1/(2g)$ in frame $M$ at the event $O$.
The event $O$ has coordinates                                                     \begin{equation}
 \zeta_0=\frac{1}{2g},~\tau_0={\frac{1}{g}}{\sinh^{-1}}(\gamma v_{ret}),~\rho=0,~\phi=0.
\label{eqn:twentyfive} \end{equation}
in frame $M$, as  can be verified from equation(\ref{eqn:twentyone}).
 It is convenient to shift the origin and define
\begin{equation}
t'=t-t_0+\frac{\gamma v_{ret}}{g}~;~z'=z-z_0+\frac{\gamma}{g}.
\label{eqn:twentysix} \end{equation}
In these coordinates, the event $O$ occurs at:
\begin{equation}
t'=\frac{\gamma v_{ret}}{g}~;~z_0'=\frac{\gamma}{g}.\label{eqn:twentyseven}
\end{equation}

Given these transformations and the form of the field in the instantaneous
Rindler frame, it is straight forward to obtain the field in the inertial
frame. Conventionally, the latter fields are expressed in terms of the
separation vector between the field point and the retarded position of the
particle. To make the comparison we will introduce
 the null vector $R^{\mu}$ with the components
\begin{equation}
R^{\mu}={x'}^{\mu}-{{x'}_0}^{\mu}=(t'-t_0',z'-z_0',\rho,\phi).
\label{eqn:twentyeight}
\end{equation}
Using the condition $R^{\mu}R_{\mu}=0$ in region $A$, we can easily show that
\begin{equation}
 \cosh           g(\tau-\tau_0)=\frac{\zeta+\zeta_0+(1/2)g{\rho}^2}{2\sqrt{\zeta}\sqrt{\zeta_0}} ~;~\zeta>0,\zeta_0=\frac{1}{2g}.\label{eqn:twentynine}
\end{equation}
Further, since the components of $v^{\mu}_{\rm ret}$ are
\begin{equation}
{v_{ret}}^0={\sqrt{2g\zeta_0}}\cosh (g\tau_0), \quad \> {v_{ret}}^z={\sqrt{2g\zeta_0}} \sinh (g\tau_0),\label{eqn:thirty}
\end{equation}
we get:
\begin{equation}
R_{\mu}{v_{ret}}^{~\mu}=2\sqrt{\zeta\zeta_0}\sinh g(\tau-\tau_0)=\sqrt{(\zeta-\zeta_0+(1/2)g{\rho}^2)^2+2g{\rho}^2\zeta_0}
\label{eqn:thirtyone}
\end{equation}
Similarly for region $B$,
\begin{equation}
\sinh g(\tau-\tau_0)=\frac{\zeta+\zeta_0+(1/2)g{\rho}^2}{2\sqrt{-\zeta}\sqrt{\zeta_0}}~;~\zeta<0,\zeta_0=\frac{1}{2g}\label{eqn:thirtytwo}
\end{equation}
and 
\begin{equation}
R_{\mu}{v_{ret}}^{~\mu}=\sqrt{(\zeta-\zeta_0+(1/2)g{\rho}^2)^2+
2g{\rho}^2\zeta_0}\label{eqn:thirtythree}
\end{equation}
which is the same as that for region $A$.

\subsection{(b) The field}
Given the field in the coaccelerating frame [equations (\ref{eqn:seventeen}) and (\ref{eqn:nineteen})] and the transformation between the inertial frame and coaccelerating frame [ (\ref{eqn:twentyone}) (\ref{eqn:twentytwo}) and  (\ref{eqn:twentysix})], we can find the field in the inertial frame. We refer to the field tensor in inertial coordinates and the electric and magnetic fields as $\it{F^{\mu\nu}}_{Min}~,~{E^i}_{Min}~{\rm and} ~ {B^i}_{Min}$ respectively. The electric field, for example, is obtained by 
\begin{equation}
{E^{z'}}_{Min}={{F_{Min}}^{z'}}_0 =\left (\frac{\partial{z'}}{\partial\tau}\right ) \left (\frac{\partial\zeta}{\partial{t'}}\right ){F^0}_{\zeta} + \left(\frac{{\partial}{z'}}{\partial\zeta} \right) \left(\frac{\partial\tau}{\partial{t'}}\right){F^{\zeta}}_0 = \frac{g}{2\zeta}({z'}^2-{t'}^2)(-\partial_{\zeta}A_0)\label{eqn:fortythree}
\end{equation}

\noindent\hbox{Therefore,}
\begin{eqnarray}
{E^{z'}}_{Min}&=&\frac{2ge\zeta_0(\zeta-\zeta_0-(1/2)g{\rho}^2)}{{\xi}^3} \nonumber \\
&=&\frac{4e}{{g}^2}\frac{({z'}^2-{t'}^2-{\rho}^2-(1/g)^2)}{[({z'}^2-{t'}^2+{\rho}^2-(1/g)^2)^2~+~4 ({\rho}^2/{g^2})]^{\frac{3}{2}}}\label{eqn:fortyfour}
\end{eqnarray}
Similarly, we obtain:
\begin{eqnarray}
{E^{\rho}}_{Min}={{F_{Min}}^{\rho}}_0=\frac{8ez'\rho}{g^2[({z'}^2-{t'}^2+{\rho}^2-(1/g)^2)^2~+~4 ({\rho}^2/{g^2})]^{\frac{3}{2}}}\label{eqn:fortyfive}\\
{B_{Min}}^{\phi}={F_{Min}}^{\rho{z}}=\frac{8et'\rho}{g^2[({z'}^2-{t'}^2+{\rho}^2-(1/g)^2)^2~+~4 ({\rho}^2/{g^2})]^{\frac{3}{2}}}\label{eqn:fortysix}
\end{eqnarray}
\begin{equation}
{E^{\phi}}_{min}={B^z}_{min}={B^{\rho}}_{min}=0.\label{eqn:fortyseven}
\end{equation}
This can be recast in a more familiar form by using equations  (\ref{eqn:thirtyone}) and (\ref{eqn:thirtythree}):
\begin{equation}
{E^{z'}}_{min}=\frac{ge({z'}^2-{t'}^2-{\rho}^2-(1/g)^2)}{2(R_{\mu}{v^{\mu}}_{ret})^3}=\frac{ge({z'}^2-{t'}^2-{\rho}^2-(1/g)^2)}{2{\gamma}^3(R-R_{z'}v_{ret})^3}
\label{eqn:fortyeight}\end{equation}
where $ R=R^0=t'-{t'}_0~;~R_{z'}=z'-{z'}_0~.$
Note that
\begin{eqnarray}
{z'}^2-{t'}^2-{\rho}^2-(1/g)^2 &=&\frac{2\gamma}{g}(R_{z'}-Rv_{ret})-2{\rho}^2\label{eqn:fortynine}\\
&=& \frac{2}{{a}_{ret}}[(1-{v_{ret}}^2)(R_z-Rv_{ret})-a_{ret}{\rho}^2]\label{eqn:fifty}
\end{eqnarray}
Therefore, we can write our answer as
\begin{equation}
{E^{z'}}_{min}=\frac{e[(1-{v_{ret}}^2)(R_{z'}-Rv_{ret})~-~a_{ret}{\rho}^2]}{(R-R_{z'}v_{ret})^3}\label{eqn:fiftyone}
\end{equation}
Similarly,
\begin{eqnarray}
{E^{\rho}}_{min}&=&\frac{e[(1-{v_{ret}}^2)\rho~+~a_{ret}R_{z'}\rho]}{(R-R_{z'}v_{ret})^3}\label{eqn:fiftytwo}\\
{B^{\phi}}_{min}&=&\frac{e[a_{ret}R\rho~+~v_{ret}(1-{v_{ret}}^2)\rho]}{(R-R_{z'}v_{ret})^3}\label{eqn:fiftythree}
\end{eqnarray}
These components  can be expressed in a more familiar vector notation as
\begin{eqnarray}
\bf E&=&\frac{e(1-{v_{ret}}^2)({\bf R}-{\bf v}_{ret} R)}{( R-{\bf R}.{\bf v}_{ret})^3}~+~\frac{e{\bf R}\times[({\bf R}-{\bf v}_{ret})\times{\bf a}_{ret}]}{(R-{\bf R}.{\bf v}_{ret})^3}\label{eqn:fiftyfour}\\
\bf B &=&\frac{{\bf R}\times{\bf E}}{R}\label{eqn:fiftyfive}
\end{eqnarray}
This is the standard result for the electromagnetic field of an arbitrary moving  charged particle (see ref[1]) 

Our results in (\ref{eqn:fiftyone}), (\ref{eqn:fiftytwo}) and (\ref{eqn:fiftythree}) have been derived for the special case of a charge in rectilinear motion. This was done to show clearly the use of our  formalism.
 In fact, one can obtain the general result quite easily. Consider the  general case, in which the motion is not restricted to a straight line. Then, one can always transform  to an {\it inertial} frame of reference, $S''$, in which the charge was at rest at the retarded event $O$. This requires us to make the usual transformation to the accelerated frame:
\begin{eqnarray}
t=t_0+\frac{\sqrt{2g\zeta}}{g}\sinh (g\tau);\qquad 
z=z_0-\frac{1}{g}+\frac{\sqrt{2g\zeta}}{g}\cosh (g\tau)\label{eqn:fiftysix}
\end{eqnarray}
followed by a Lorentz transformation to bring ${\bf v}_{\rm {ret}}$
to zero. Working in a similar fashion, we will land up with simpler expressions for the fields:
\begin{equation}
{\bf E}=\frac{e{\bf R}}{R^3}~+~\frac{{\bf R} \times({\bf R} \times {\bf a}_{ret})}{R^3}; \qquad {\bf B}=\frac{{\bf R} \times {\bf E}} R\label{eqn:fiftyeight}
\end{equation}
This gives the field in the Lorentz frame in which the charge has zero velocity at the retarded event. By making a Lorentz transformation with an arbitrary velocity ${\bf v}$, we can get fields in equations (\ref{eqn:fiftyfour}), (\ref{eqn:fiftyfive}). More formally, one can show that the fields in (\ref{eqn:fiftyeight}) can be obtained from the following Lorentz invariant expression:
\begin{equation}
F_{\mu\nu}=\frac{e}{(R^{\sigma}v_{\sigma})^3}[(R_{\mu}v_{\nu}-R_{\nu}v_{\mu})+R_{\nu}R^{\sigma}(v_{\mu}a_{\sigma}-v_{\sigma}a_{\mu})-R_{\mu}R^{\sigma}(v_{\nu}a_{\sigma}-v_{\sigma}a_{\nu})]\label{eqn:sixty}
\end{equation}
Then, since this is a tensor equation under Lorentz transformations, it will give the fields in a frame $S$ in which the retarded velocity is arbitrary. A simple calculation shows that (\ref{eqn:sixty}) reduces to (\ref{eqn:fiftyfour}) and (\ref{eqn:fiftyfive}) in this case. (This form is derived in ref [3] in a very complicated manner. A simple  derivation of (\ref{eqn:sixty}) is given in Appendix C).

\section{Electromagnetic fields in the infinitesimal neighbourhood of an accelerating charge}
As before, let us consider a rectilinearly moving charge with arbitrary velocity and acceleration along the $z$-axis of frame $S$. The instantaneous, uniformly accelerating frame is $M$. At the event $O$, the charge is at rest in $M$ and with zero acceleration at the point $\it \zeta=\zeta_0$. We are interested in calculating the finite part of the force exerted on the charge by it's own field at the event $O$. It is now convenient to use the  coordinate Z introduced earlier with the definition :
\begin{equation}
Z=\frac{\sqrt{2g\zeta}}{g}\label{eqn:sixtyone}
\end{equation}
In the frame with coordinates $\it(\tau,Z,\rho,\phi)$, the charge, is at rest in $M$ at $ Z=Z_0=(1/g)$ at event $O$. In these coordinates, the fields corresponding to the retarded event $O$ are :

\begin{equation}
E^Z=\frac{4e(Z^2-{Z_0}^2-{\rho}^2)}{g^2[(Z^2-{Z_0}^2+{\rho}^2)^2+4{\rho}^2{Z_0}^2]^{\frac{3}{2}}}; \qquad E^{\rho} =  \frac {8 e\rho Z} {g^2[(Z^2-{Z_0}^2+{\rho}^2)^2+4{\rho}^2{Z_0}^2]^{\frac{3}{2}}}
\label{eqn:sixtytwo}
\end{equation}
These  are same as those given by equations (\ref{eqn:fortyfour}) and (\ref{eqn:fortyfive}), but expressed  in the new coordinate $ Z$. 

We will analyse the  situation  described in figure 2. The  event $O$,  at which   the charge has zero velocity and acceleration, corresponds to  $\tau=0.$ Consider now an event $O'$ along the worldline of the charge.
 The forward light ray travelling from this event is seen to intersect the $ Z$
axis at the point $P$ , which is an event  simultaneous with event $O$. We want to study the fields at $P$ in the limit $P \rightarrow O$. In this limit, $O'\rightarrow O$
and the fields at $P$ are those due to the charge in it's own infinitesimal neighbourhood. (Since the metric is noninertial, the curve $O'P$ will not
be a straight line. But we are only interested in the limit of $O'\to O$
when the actual form of the curve is irrelevant).

The field due to the {\it retarded point} $O$ is a  static field, given by equation (\ref{eqn:sixtytwo}). However, we are interested in calculating the field at event $P$ due to the {\it event} $O'$. At $O'$, the charge is not at rest in frame $M$. So, the field given by equation (\ref{eqn:sixtytwo}) will not work. But, in the limit $O$ $\rightarrow$ $O'$, we can make a correction to the field given by equation (\ref{eqn:sixtytwo}), which can account for it's motion at event $O'$. To illustrate this, let us approach the infinitesimal neighbourhood of the charge along the $ Z$ axis. If we ignore for the moment the motion of the charge then the field along the $ Z$ axis at event $P$ (in the limit $O$ $\rightarrow$ $O'$) is:
\begin{equation}
E^Z=\frac{4e}{g^2}\frac{1}{(Z^2-{Z_0}^2)^2}; \qquad E^{\rho}=0 \label{eqn:sixtythree}
\end{equation}
Let the event $O'$ occur at $ \tau=-{\tau}_0 $. In the approximation that the charge was at rest at $O'$, it can be verified that 
\begin{equation}
R_{\mu}{v^{\mu}}_{ret}=\frac{g}{2}(Z^2-{Z_0}^2)\label{eqn:sixtyfive}
\end{equation}
which can be obtained from equations (\ref{eqn:thirtyone})  and (\ref{eqn:sixtyone}) after setting $ \rho=0$.
So the field can be expressed as:
\begin{equation}
E^Z=\frac{e}{(R_{\mu}{v_{ret}}^{\mu})^2}\label{eqn:sixtysix}
\end{equation}

Let us next account for the charge's motion at $O'$. We begin by noting that, in arriving at  equation  (\ref{eqn:thirtyone}), we used the fact that the charge was at rest at the retarded event.
 When we take into account the motion of the charge at $O'$, the expression for $(R_{\mu} v^{\mu}_{ret}$) will be modified. Since the Coulomb part of the field dominates as $O' \rightarrow O$, the leading term in the field is still given by (\ref{eqn:sixtysix}) with the corrected expression for  $ R_{\mu}{v_{ret}}^{\mu}$.
 If  velocity of the charge at $O'$ is $ {u^Z}(-{\tau}_0)$, then --- in the limit we are interested in ---  the updated $ (R_{\mu}{v_{ret}}^{\mu})$ is:
\begin{equation}
R_{\mu}{v_{ret}}^{\mu}=\frac{g}{2}(Z^2-{Z_0}^2(-{\tau}_0))~-~(Z-{Z_0}(-{\tau}_0)){u^Z}(-{\tau}_0)\label{eqn:sixtyseven}
\end{equation}
So the expression for the field which accounts for the motion of the charge at $O'$ will be:
\begin{equation}
E^Z (P) =\frac{4e}{g^2[Z^2-{Z_0}^2(-{\tau}_0)-\frac{2}{g}(~Z-Z_0(-{\tau}_0)~){u^Z}(-{\tau}_0)]^2}\label{eqn:sixtyeight}
\end{equation}
At event $O$, the velocity and acceleration of the charge are zero in $M$. However, the rate of change of acceleration is nonzero, and we denote this quantity by $  \dot{\alpha} $. Then, all the relevant quantities  at time $ \tau=-{\tau}_0 $ can be expressed in terms of $ \dot{\alpha}$ alone, in the limit $O$ $\rightarrow$ $O'$ (that is when $ {\tau}_0 \rightarrow 0$). Making a Taylor expansion about $ \tau=0$ , we get:
\begin{equation}
Z_0(-{\tau_0})\cong Z_0-\frac{{\tau_0}^3}{6}\dot{\alpha}
\label{eqn:sixtynine}
\end{equation}
 \hbox{and} 
\begin{equation}
u^Z(-{\tau}_0)\cong\frac{{\tau_0}^2}{2}\dot{\alpha}\label{eqn:seventy}
\end{equation}
where $ Z_0=Z_0(0)=g^{-1} $ .
(Here, only terms upto order ${\tau_0}^3$ need to be retained, in the limit of $\tau_0 \rightarrow$ 0). In this limit, writing $ Z=Z_0+\delta$, we find that
\begin{equation}
{E^Z}(P)\cong\frac{4e}{g^2[{\delta}^2+2Z_0 {\delta}+(Z_0{\tau_0}^3\dot{\alpha}/3)-Z_0\delta{\tau_0}^2\dot{\alpha}]^2}
\label{eqn:seventyone}\end{equation}
From equation (\ref{eqn:twentynine}), we get, after setting $\tau= 0$, $\rho = 0$ and replacing ${\tau_0}$ by $ {-\tau_0}$
\begin{equation}
\sinh (g\tau_0)=\frac{\zeta-\zeta_0}{2\sqrt{\zeta\zeta_0}}\cong\frac{\delta}{Z_0}
\label{eqn:seventytwo}\end{equation}
in the limit $\delta\rightarrow$0. Hence,  in the limit $\tau_0\rightarrow 0$:
\begin{equation}
\tau_0\cong\delta\label{eqn:seventythree}
\end{equation}
We  are interested in  the force exerted on the charge by its own field, which is equal to $ eE^Z(P)$ in the limit $\delta\rightarrow$0.
This force is given by: 
\begin{equation}
F^Z= e E^Z(P)\cong \frac{e^2}{{\delta}^2}-\frac{e^2g}{\delta}+\frac{2e^2}{3}\dot{\alpha}+\frac{3e^2g^2}{4}\label{eqn:seventyfour}
\end{equation}
where we have  expanded the expression for $\it E^Z $ 
in the binomial series  in the infinitesimal parameter $\it\delta$. It is understood that we should evaluate this expression in the limit of $\it\delta\rightarrow$ 0 .

The first two terms diverge as $\it\delta\rightarrow$ $0$. This point is extensively discussed in  literature, and these terms arise from the self energy of a charged particle due to interaction with its own electromagnetic field and  are expected to be absorbed by mass-renormalization.
      There is also the constant (last) term, which is uninteresting, since the first two terms are already divergent. We would have
landed up with these three terms, even if we had not accounted for the motion of the charge at event $O'$. 
It is the third term, which has the derivative 
of the acceleration, which is the most interesting term. We have been able to obtain  it because we accounted for the motion of the charge at event $O'$. {\it It is this term which accounts for the effect of radiation reaction on the charge.}

All
these terms were first found by Dirac for arbitrary motion of the charge in an
inertial frame. The general expression in the inertial frame obtained by Dirac is [ref. 4; also see pp 141-144 of ref.3]
\begin{equation}
f^{\mu}\cong \frac{e^2}{\sqrt{1-(\delta) a^{\lambda}u_{\lambda}}}[\frac{u^{\mu}}{{\delta}^2}-\frac{a^{\mu}}{2\delta}-\frac{a^{\mu}~(a^{\lambda}u_{\lambda})}{2}+\frac{ga^{\mu}}{8}-\frac{u^{\mu}~(\dot{a_{\lambda}}v^{\lambda})}{2}+\frac{2}{3}(\dot{a^{\mu}}-v^{\mu}(\dot{a_{\lambda}}v^{\lambda})~)]\label{eqn:seventyfive}
\end{equation}
Here, $\it a^{\mu}$ is the 4-acceleration, $\it v^{\mu}$ the 4-velocity, $\it \delta=R^{\lambda}v_{\lambda}$ and $\it u^{\mu}\equiv[({R^{\mu}-\delta  v^{\mu}})/({\delta})]$. These are the leading  terms in an expansion in $\it \delta$  in the limit $\delta  \rightarrow 0$.

Computing the above  expression for  $\it f^{\mu}$ in the inertial frame is a laborious task. Our formalism makes the corresponding computation  very simple. In fact, we  calculated it the same way we calculated the fields: We transform to an inertial frame $S''$, in which the charge is at rest at the retarded event and  find the expression for the force in the comoving accelerating frame, which is given by equation  (\ref{eqn:seventyfour}). By transforming back to $S''$ and making an arbitrary  Lorentz transformation to $S$ gives the force (\ref {eqn:seventyfive}) in the general inertial frame starting from (\ref{eqn:seventyfour}). (We give the explicit  procedure in  Appendix B.) Here we shall transform only the radiation reaction term, and show that it gives the correct radiation reaction in the inertial frame, for the case of rectilinear motion.

Let $\it f^z$ denote the radiation reaction force in the inertial frame $S$.
 Then, if $\it {F_{rad}}^Z $denote
the radiation reaction force in frame $M$, then  using the  equations (\ref{eqn:twentyone}),  (\ref{eqn:twentyfive}) and (\ref{eqn:sixtyone}), we have 
\begin{equation}
f^z=\left(\frac{\partial z}{\partial Z}\right)_{ret} ~{F_{rad}}^Z=\frac{2e^2}{3}~\dot{\alpha}~\cosh (g\tau_0)\label{eqn:seventysix}
\end{equation}
The derivative of the retarded  acceleration $\it {\dot a}^{\mu}_{ret}=da^{\mu}/(d\tau),$ as measured in the frame $S$ is related to $\it \dot{\alpha}$ by:

\begin{eqnarray}
\dot{a^0}&=&g^2~\cosh (g\tau_0)+\dot{\alpha}~\sinh (g\tau_0)\label{eqn:seventyseven}\\
\dot{a^z}&=&g^2~\sinh (g\tau_0)+\dot{\alpha}~\cosh (g\tau_0)\label{eqn:seventyeight}\end{eqnarray}
Then, 
 using equation (\ref{eqn:seventysix}), we get
\begin{equation}
f^z=\frac{2e^2}{3}~\dot{\alpha}~\cosh (g\tau_0)=\frac{2e^2}{3}(\dot{a^z}-v^z(\dot{a^\mu}v_{\mu})~)\label{eqn:seventynine}
\end{equation}
which is indeed the radiation reaction force in the inertial frame $S$.
For an arbitrarily moving charge, one can first transform to the instantaneous  Lorentz frame, in which (putting $\it v^z$=0) :
\begin{equation}
f^z=\frac{2e^2}{3}\dot{a^z}\label{eqn:eighty}
\end{equation}
and  make a Lorentz transformation to an arbitrary inertial frame, to get~:
\begin{equation}
f^{\mu}=\frac{2e^2}{3}(\dot{a^{\mu}}-v^{\mu}(\dot{a^{\nu}}v_{\nu})~)
\label{eqn:eightyone}\end{equation}
which is the correct expression for radiation reaction. 

An attempt is made in ref [2] to relate the {\it radiated power} to the force acting between the charge and the ficticious charge density at the horizon. Our result is more general and gives the actual {\it radiation reaction force} itself. Further we did not have to use the {\it ficticious}, coordinate dependent, charge density to interpret a {\it real} effect.

\section {Conclusions:}

The radiation of electromagnetic waves by a charged particle and 
the consequent radiation reaction force are issues of considerable 
theoretical significance and have attracted the attention of researchers
over decades. We believe that the approach outlined in this paper
throws light on these processes and clarifies the conceptual issues
involved in the problem. To begin with, we have been able to derive
the radiation field as arising out of a static Coulomb field in a non 
inertial frame through a general coordinate transformation. This is of 
some conceptual importance since one believes that physics should be 
independent of the coordinate system. Secondly, we have been able to show 
that the key contribution to radiation reaction arises because of the 
deviation of the trajectory from that of a uniformly accelerated one.
This deviation, which essentially modifies the expression $R^\mu v_\mu$, 
has a purely geometrical origin in the locally co-accelerating frame.
Since the lowest order deviation will be proportional to the rate of 
change of acceleration, it is clear that the radiation reaction force
should be proportional to the same; that is, it should be proportional 
to the third derivative of the trajectory.

It will be of interest to see whether these results allow us to tackle 
the question of self-force in curved space time and to generalize the 
various expressions to an arbitrary curved background. We hope to address
these questions in a future publication.

\section{APPENDIX }

\subsection{Solution to Maxwell's Equations for a charge at rest in a uniformly accelerating frame}

The scalar potential $ A_0$ due to a charge at rest in a uniformly accelerating frame satisfies the following equation [see equation (\ref{eqn:sixteen})]:
\begin{equation}
\rho{\partial_{\zeta}}^2A_0+\frac{1}{2g\zeta}\partial_{\rho}(\rho\partial_{\rho}A_0)=-4\pi e\delta(\zeta-\zeta_0)~\delta(\rho)~\delta(\phi)\label{eqn:eightytwo}
\end{equation}
We shall find a solution to this equation by studying  a different, but related problem.

Consider the problem of a charge placed at rest outside the horizon of a Schwarzschild black hole. 
In the  spherical polar coordinates $\it (r,\theta,\phi)$, the charge is placed at $\it r=r',~\theta=0$ .
 The metric is given by:
\begin{equation}
ds^2=(1-\frac{2M}{r})dt^2-\frac{dr^2}{(1-\frac{2M}{r})}-r^2(d{\theta}^2+{\sin }^2\theta d\phi^2)\label{eqn:eightythree}
\end{equation}
Maxwell's equations are seperable in these coordinates.
The differential equation satisfied by $\it A_0$ in this metric is:
\begin{equation}
\frac{1}{r^2}\partial_r(r^2\partial_rA_0)+\frac{1}{(1-\frac{2M}{r})}\frac{1}{r^2 \sin \theta}\partial_{\theta}(\sin {\theta}\partial_{\theta}A_0)=-4\pi e\delta(r-r')\delta(cos{\theta}-1)\label{eqn:eightyfour}
\end{equation}
The solution to this equation can be expressed in a closed form (see ref. 5):
\begin{equation}
A_0=\frac{e[(r-M)(r'-M)-M^2cos\theta]}{rr'\sqrt{(r-M)^2+(r'-M)^2-2(r-M)(r'-M)\cos \theta-M^2{\sin }^2\theta}}\label{eqn:eightyfive}\end{equation}

Now, if the charge is placed infinitesimally close to the horizon and all measurements made in an arbitrarily small region around the charge,  then the horizon would appear ``flat''.
Mathematically, if we introduce a coordinate $\it \zeta$ by:
\begin{equation}
r=2M+\zeta~;~2M\gg\zeta\label{eqn:eightysix}
\end{equation}
then we can write
\begin{equation}
(1-\frac{2M}{r})\cong\frac{\zeta}{2M}=2g\zeta\label{eqn:eightyseven}
\end{equation}
where $\it g=(1/4M)$ is the effective surface gravity of the horizon.
If we fix our origin at $\it \theta=0,~r\cong2M$ and restrict all observations perpendicular to the z-axis to a very small region, then $\it \rho\cong2M\sin \theta\cong2M\theta$. 
In this limit, 
\begin{equation}
r^2(d{\theta}^2+{\sin}^2\theta~d{\phi}^2)\cong (2M)^2(d{\theta}^2+{\theta}^2~d{\phi}^2)\cong d{\rho}^2+{\rho}^2~d{\phi}^2
\label{eqn:eightyeight}\end{equation}
giving, 
\begin{equation}
ds^2=2g\zeta~dt^2-\frac{d{\zeta^2}}{2g\zeta}-d\rho^2-{\rho}^2~d{\phi}^2
\label{eqn:eightynine}\end{equation}
which is identical to the metric given by equation (\ref{eqn:seven}) .

In this approximation, we are neglecting curvature of spacetime such that the horizon appears like a plane $(\zeta=0)$ as in a uniformly accelerating frame. More importantly, it gives us an ansatz to find a solution to equation (\ref{eqn:sixteen}), 
by using this approximation in the expression for $ A_0$ given by equation (\ref{eqn:eightyfive}) .
Straightforward algebra gives:
\begin{equation}
A_0 \cong \frac{ge(\zeta+\zeta_0+\frac{1}{2}g\rho^2)}{\sqrt{(\zeta-\zeta_0+\frac{1}{2}g
\rho^2)^2 + 2g \rho^2 \zeta_0}}
\end{equation}
which is, an exact solution to equation (\ref{eqn:eightytwo}).

\noindent
\subsection {  The Lorentz-Dirac formula}
\noindent
Consider a charge moving in an arbitrary trajectory in an inertial frame $S$.  As before, we construct a comoving, uniformly accelerating frame $M$, in which the expression for the self-force is given by (\ref{eqn:seventyfour}). Consider now the expression (\ref{eqn:seventyfive}) in a frame in which
\begin{equation}
v^{\mu} = (1,0,0,0), \quad \dot{a^{\mu}}=(g^2,\dot{\alpha},0,0), ~a^{\mu}=(0,g,0,0).
\end{equation}
The expression reduces to
\begin{equation}
f^{\mu}\cong e^2~a^{\mu}(\frac{1}{g{\delta}^2}- \frac{1}{\delta}+\frac{3g}{4}) +\frac{2e^2}{3}[\dot{a^{\mu}}-v^{\mu}(\dot{a_{\lambda}}v^{\lambda})]
\end{equation}
so that 
\begin{equation}
f^z\cong e^2(\frac{1}{\delta^2}-\frac{g}{\delta}+\frac{3g^2}{4}+\frac{2e^2}{3}\dot{a^z})
\label{eqn:eightyfive}\end{equation}
(It is assumed that $\delta \rightarrow 0$ limit is considered; also note that $u^{\mu} \rightarrow a^{\mu}$ in this limit.) This is identical to (\ref{eqn:seventyfour}) when we use (\ref{eqn:seventyeight}) with $\tau_0=0$. If one wants the expression for $\it f^{\mu}$ in the frame $S$, all one has to do is find the expression in $M$ (which is a simpler task compared to  directly calculating  it in the frame $S$, as is normally done), given by equation (\ref{eqn:seventyfour}), transform it to the inertial frame (to get expression (\ref{eqn:eightyfive})), and finally to transform it to $S$ by making a Lorentz transformation. This leads to the  Lorentz-Dirac expression, given by equation (\ref{eqn:seventyfive}).

\noindent
\subsection{ Covariant form of the field of an arbitrarily moving charge:}

\noindent
It is possible to express the field tensor $F^{\mu \nu}$ produced by an arbitrarily moving charge, ina manifestedly Lorentz invariant form. Though the result is obtained in ref.3, the derivation is quite cumbersome. We give here a clearer and simpler derivation of this formula. Maxwell's equations in inertial coordinates in flat spacetime are:
\begin{equation}
\partial_{\mu}F^{~\mu\nu}=4\pi j^{\nu};
F_{~\mu\nu}=\partial_{\mu}A_{\nu}-\partial_{\nu}A_{\mu}
\end{equation}
Combined together in the Lorentz gauge ($\it \partial^{\mu} A_{\mu}$)=0,
 we have:
\begin{equation}
\Box A_{\mu}=4\pi j_{\mu}
\end{equation}
which has the solution:
\begin{equation}
A_{\mu}(x)=4\pi\int~d^4 x~G_{ret}(x-y)~j_{\mu}(y)
\end{equation}
where $\it G_{ret}$ is the retarded Green's function, satisfying:
\begin{equation}
\Box G_{ret}(x-y)=\delta^4 (x-y)~;~G_{ret}(x-y)=0 ~for~x^0 < y^0~.
\end{equation}
The current $\it j^{\mu}(x)$ for a point charge moving along a worldline $\it z^{\mu}=z^{\mu}(\tau)$ with a 4-velocity $\it u^{\mu}(\tau)$ is given by:
\begin{equation}
j^{\mu}(x)=e\int~d\tau~\delta^4 [x-z(\tau)]~u^{\mu}(\tau)
\end{equation}
so that
\begin{equation}
A_{\mu}(x)=4\pi e\int~d\tau~G_{ret}[x-z(\tau)]u_{\mu}(\tau)
\end{equation}
Now, let $\it R^{\mu}=x^{\mu}-z^{\mu}(\tau)$ . Then, 
\begin{equation}
G_{ret}[x-z(\tau)]=\frac{1}{2\pi}\delta(s^2)~\theta(x^0-z^0)~;~s^2\equiv R^{\mu}R_{\mu}
\end{equation}
giving,
\begin{equation}
A_{\mu}(x)=2e\int~d\tau~\delta(s^2)~u_{\mu}(\tau)
\end{equation}
and
\begin{equation}
\partial_{\nu}A_{\mu}(x)=2e\int~d\tau~\partial_{\nu}\delta(s^2)~u_{\mu}(\tau)=2e\int~d\tau~\frac{d\delta(s^2)}{ds^2}\frac{\partial s^2}{\partial x^{\nu}}~u_{\mu}(\tau)
\end{equation}
Now,
\begin{equation}
\frac{\partial s^2}{\partial x^{\nu}}=2R_{\nu}
\end{equation}
Therefore,
\begin{equation}
\partial_{\nu}A_{\mu}(x)=4e\int~d\tau~\frac{d\delta(s^2)}{d\tau}~(\frac{ds^2}{d\tau})^{-1}~R_{\nu}u_{\mu}(\tau)
\end{equation}
Also, 
\begin{equation}
\frac{ds^2}{d\tau}=-2\rho~;~\rho\equiv R^{\mu}u_{\mu}
\end{equation}
leading to
\begin{eqnarray}
\partial_{\nu}A_{\mu}(x)&=&-2e\int~d\tau~\frac{d\delta(s^2)}{d\tau}~(\frac{R_{\nu}u_{\mu}}{\rho})\nonumber\\
&=&2e\int~d\tau~\delta(s^2)~\frac{d}{d\tau}(\frac{R_{\nu}u_{\mu}}{\rho})\nonumber\\
&=&\frac{e}{\rho}~\frac{d}{d\tau}(\frac{R_{\nu}u_{\mu}}{\rho})\mid_{ret}
\end{eqnarray}
It follows that:,
\begin{equation}
F_{\mu\nu}=\partial_{\mu}A_{\nu}-\partial{\nu}A_{\mu}=\frac{e}{\rho}~\frac{d}{d\tau}(\frac{R_{\mu}u_{\nu}-R_{\nu}u_{\mu}}{\rho})\mid_{ret}
\end{equation}
Differentiating the expression, we get
\begin{equation}
F_{\mu\nu}=\frac{e}{\rho^3}[(R^{\sigma}u_{\sigma})(R_{\mu}a_{\nu}-R_{\nu}a_{\mu})~+~(1-R^{\sigma}a_{\sigma})(R_{\mu}u_{\nu}-R_{\nu}u_{\mu})]
\end{equation}
Using $R^{\sigma}R_{\sigma}=0~;~R^{\mu}=(R,{\bf R})~;~u^{\mu}=\gamma (1,{\bf u})~;~a^{\mu}=\gamma(\dot\gamma,{\bf u}{\dot\gamma}+\gamma {\bf a})$, we get the electric and magnetic fields as :
\begin{eqnarray}
{\bf E}&=&\frac{e(1-u^2)({\bf R}-{\bf u} R)}{(R-{\bf R} . {\bf u})^3}~+~\frac{e{\bf R}\times[({\bf R}-{\bf u } R)\times {\bf a}]}{(R-{\bf R} . {\bf u})^3}\\
{\bf B}&=&\frac{{\bf R}\times {\bf E}}{R}
\end{eqnarray}
These are the standard text book expressions for the fields.


\begin{thebibliography}{20}
\bibitem{ll72}
L.~D.~ Landau and E.~M.~ Lifschitz, (1972) Classical theory of fields, (Pergammon)
\bibitem{rohr94} 
F.~J.~ Alexander and U.~H.~ Gerlach (1991) Phys. Rev. D., {\bf 44}, 3887.
\bibitem{ag91}
F. Rohrlich, (1964) Classical charged particle, (Addison-Wesley)
\bibitem{dirac38}
P.A.M. Dirac (1938) Proc. Roy. Soc. (London) A {\bf 165}, 199
\bibitem{mem86}
 K.S. Thorne, R.H. Price, 
D. A. Macdonald, (Eds) (1986) Black Holes: The Membrane Paradigm. (Yale Univ. Press) 

\end{thebibliography}
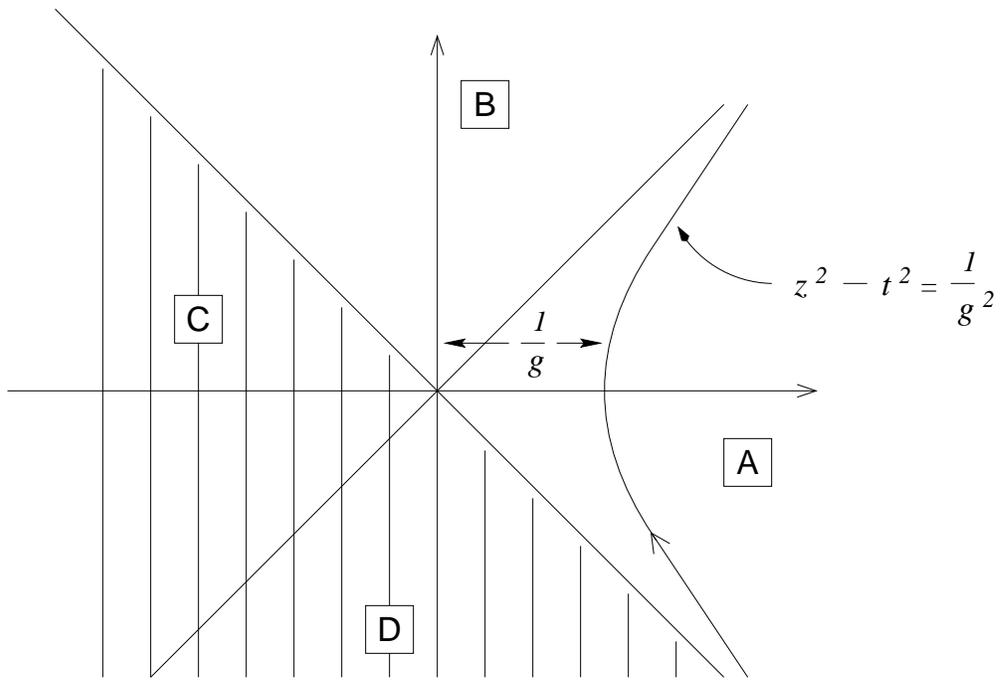
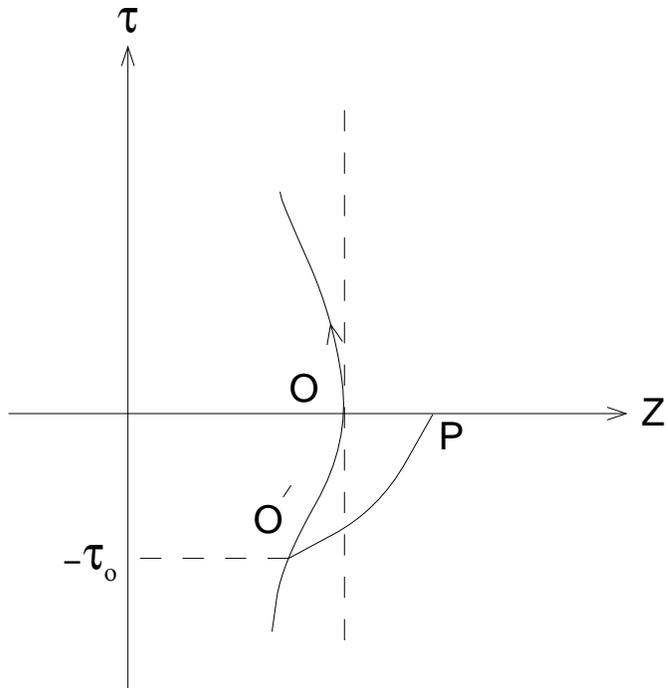
\end{document}